\documentclass[prb,twocolumn,floatfix]{revtex4-1}
\usepackage{amsmath,amsfonts}
\usepackage{graphicx}
\usepackage{dcolumn}
\usepackage{color}
\usepackage[
breaklinks,
colorlinks,
linkcolor=blue,
citecolor=blue,
urlcolor=blue]{hyperref}
\newcommand{\tr}{\mbox{Tr}}

\begin{document}

\author{Alexei A. Kananenka}
\author{Jordan J. Phillips}
\author{Dominika Zgid}
\affiliation{Department of Chemistry, University of Michigan,
Ann Arbor, Michigan 48109, United States}


\title{Efficient temperature-dependent Green's functions methods for
realistic systems: compact grids for orthogonal polynomial transforms}

\begin{abstract}
   The temperature-dependent Matsubara Green's function that is used
   to describe temperature-dependent behavior is expressed on a
   numerical grid. While such a grid usually has a couple of hundred
   points for low-energy model systems, for realistic systems in 
   large basis sets the size of an accurate grid
   can be tens of thousands of points, constituting a severe
   computational and memory bottleneck. In this paper, we determine 
   efficient imaginary time grids for the temperature-dependent
   Matsubara Green's function formalism that can be used for
   calculations on realistic systems. We show that due to the use
   of orthogonal polynomial transform, we can restrict the imaginary time
   grid to few hundred points and reach micro-Hartree accuracy in
   the electronic energy evaluation. Moreover, we show that only a
   limited number of orthogonal polynomial expansion coefficients 
   are necessary to preserve accuracy when working with a dual 
   representation of Green's function or self-energy and transforming 
   between the imaginary time and Matsubara frequency domain.  
\end{abstract}

\maketitle


\section{Introduction}

The use of numerical grids in calculations for realistic systems has
a long history in quantum chemistry simulations. For example, in
density functional theory (DFT) a numerical integration is necessary 
for the evaluation of the exchange-correlation contribution to the 
density functional~\cite{:/content/aip/journal/jcp/88/4/10.1063/1.454033,doi:10.1080/00268979300100651,GILL1993506,:/content/aip/journal/jcp/104/24/10.1063/1.471749,JCC:JCC10211}. 
Similarly, in Laplace transformed MP2 (LT-MP2) a quadrature is used 
to represent an integral that leads to the removal of the energy 
denominators~\cite{Almlof1991319,:/content/aip/journal/jcp/96/1/10.1063/1.462485,B803274M,B804110E,B802993H,:/content/aip/journal/jcp/110/8/10.1063/1.478256}. Recently, an implementation of random phase approximation 
(RPA)~\cite{doi:10.1021/ct5001268} appeared that uses an efficient 
imaginary time grid to yield temperature-independent RPA energy. 
The above mentioned methods are just a few examples of using 
efficient quadrature, a more extensive literature on the subject can 
be found in refs~\citenum{C5CP03410H,C5CP01214G,C5CP01090J,C5CP02561C,C5CP01183C,C5CP00995B,C5CP01093D,C5CP01215E,C5CP01173F,C5CP01222H,C5CP01211B,C5CP00437C,C5CP00110B,C5CP00351B,C4CP05821F,C5CP00333D,C5CP00934K,C5CP00345H,C5CP00320B,C5CP00352K}.
Thus, it is fair to say that extensive knowledge exist on representing 
temperature-independent quantities on a grid when ground state methods 
are used. 
However, very little is known about how to efficiently represent 
temperature-dependent data on finite-temperature imaginary axis 
Matsubara grids.

Several factors distinguish the finite temperature Green's function 
from the zero-temperature Green's function formalism.
Firstly, let us note that the temperature-dependent Green's function 
is a discrete object for which the grid points $i\omega_n$ are 
spaced according to the Matsubara grid $w_n=(2n+1)\pi/\beta$, 
where $n \in \mathbb{Z}$ and $\beta=1/k_\text{B}T$ is the inverse temperature. 
In comparison, the zero-temperature Green's function represented 
on imaginary axis is a continuous function. Similarly, the temperature-dependent 
imaginary time Green's function is an antiperiodic function 
between $0$ and $\beta$, while the zero-temperature Green's function 
of imaginary time is a non-periodic function decaying rapidly and 
smoothly to zero. Consequently, traditional quadratures developed 
for zero-temperature RPA Green's functions used in ref~\citenum{doi:10.1021/ct5001268} 
or LT-MP2 to represent 
denominators~\cite{Almlof1991319} are not suitable for temperature-dependent 
Green's function calculations. 

Currently, temperature-dependent Green's function calculations 
are mostly done for low-energy model systems. While such calculations 
for large realistic systems are still in their infancy, one can 
easily imagine that they could be very important in materials 
science for materials with small band gaps where the change of 
properties with temperature is significant and multiple states 
can be easily populated, or for a system which exhibit 
temperature-dependent phase transition caused by the electronic 
degrees of freedom.

\begin{figure}[htbp]
\begin{center}
\includegraphics[width=0.49\textwidth]{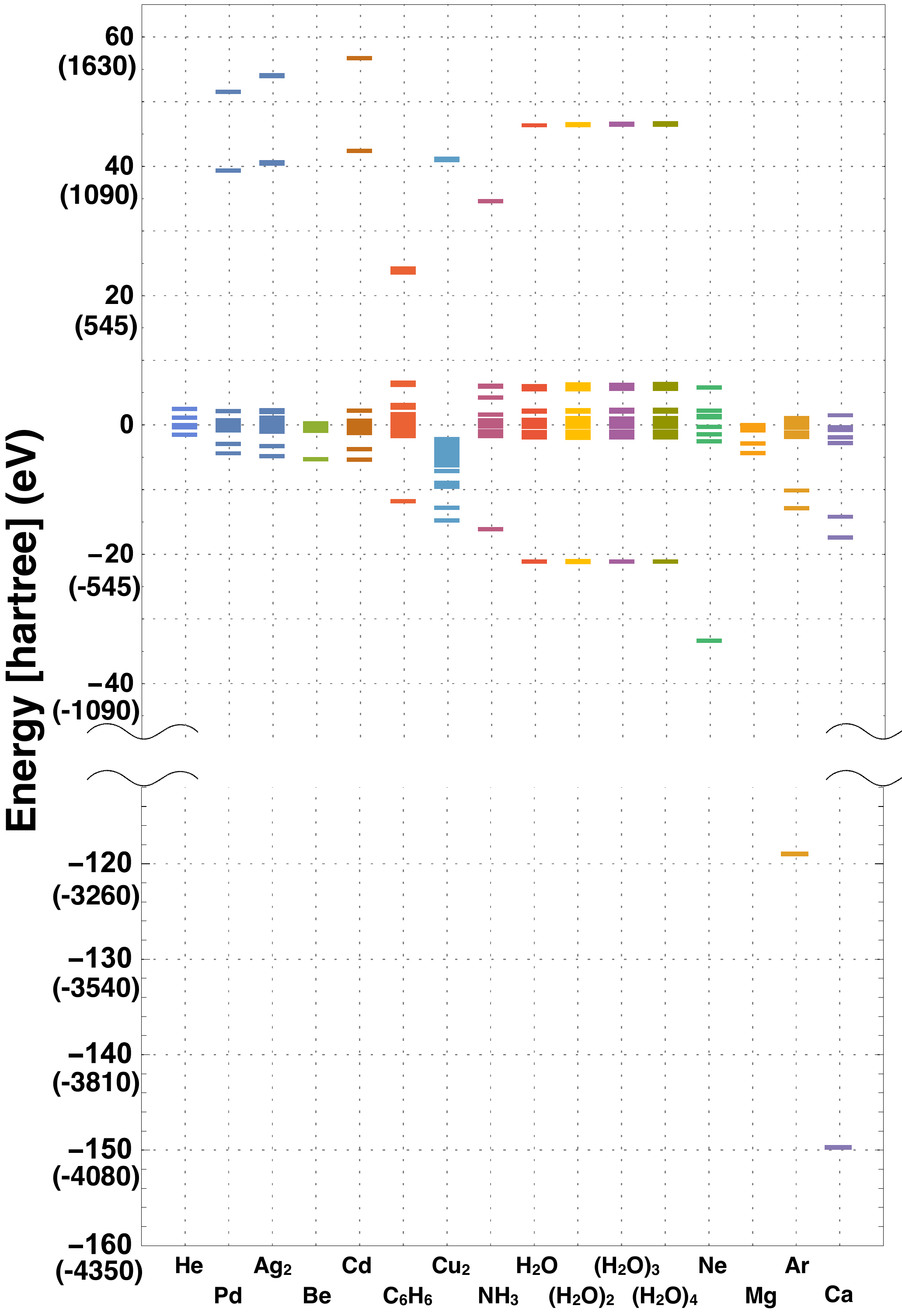}
\caption{Orbital energies for atoms and molecules in large basis sets containing diffuse orbitals.}
\label{fig:energy_window}
\end{center}
\end{figure}

Unlike the low-energy models, the orbital energies in realistic 
systems span a huge energy window frequently varying even between 
$-3000$ eV to $300$ eV, see Figure~\ref{fig:energy_window}. 
Thus, when quantitative accuracy in the calculations of realistic 
systems is desired, new challenges arise that are not present 
in model system calculations. 

The temperature-dependent Green's functions have to be represented 
on a numerical grid with a spacing defined by the temperature but 
covering the energy window spanned by the realistic system. 
This usually results in a grid containing thousands of frequencies.
Consequently, the Fourier transform from the imaginary time to the 
imaginary frequency axis, according to Nyquist theorem, requires 
twice as many imaginary time points as frequency points to yield 
accurate results. Having several thousands data points makes the 
calculations for realistic systems extremely challenging 
even if each grid point can be calculated in parallel. 

Motivated by the aforementioned challenges, we first determine how 
the imaginary time grid can be truncated to a reasonable size and 
how the resulting errors can be controlled. To achieve it, we replace 
the numerical Fourier transform ($i\tau \rightarrow i\omega$) with an 
orthogonal polynomial transform ($i\tau \rightarrow L$) and 
($L \rightarrow i\omega$), where as $L$ we denote expansion 
coefficients of a Green's function or self-energy in an orthogonal 
polynomial basis.
Subsequently, we examine if 
{\bf (i)} 
the expansion coefficients $L$ can be produced 
using a smaller number of grid points than currently used in the 
numerical Fourier transform, and if 
{\bf (ii)} the number of expansion coefficients $L$ is small 
enough that they can be easily stored in realistic calculations.

This paper is organized as follows. In section~\ref{theory}, we 
review the necessary theoretical background leading to compact 
imaginary time grids and orthogonal polynomial transforms 
($i\tau \rightarrow L \rightarrow i\omega$) using Legendre expansion 
coefficients. In section~\ref{GF2}, 
we discuss how an orthogonal polynomial transform can be used 
in the second order iterative Green's function method (GF2) to 
reduce the size of the imaginary time grid. In section~\ref{results}, 
we present numerical results showing that the required number of 
imaginary time points is much smaller than in the original uniform and 
power-law grid and that the number of required expansion 
coefficients can be kept small even if the micro-Hartree ($\mu$Ha) 
accuracy is desired. Finally, we present conclusions in 
section~\ref{conclusions}.

\section{Theory}\label{theory}

In many Green's function methods two Green's function
($G$) or self-energy ($\Sigma$) representations are used: 
an imaginary time representation $G(i\tau)$ or $\Sigma(i\tau)$ 
and imaginary frequency representation $G(i\omega)$ or $\Sigma(i\omega)$.
In an efficient implementation, one frequently changes from 
$i\tau$ to $i\omega$ and back, depending if it is more 
computationally advantageous to work with a $i\tau$ or $i\omega$ 
representation. 
Note that while the above statement is general and the $i\tau$ 
to $i\omega$ transform may be present in temperature-independent 
calculations such as RPA or GW~\cite{doi:10.1021/ct5001268,PhysRevLett.74.1827,:/content/aip/journal/jcp/135/7/10.1063/1.3624731}, in this 
paper we focus exclusively on the imaginary time and frequency 
used for temperature-dependent Green's 
functions~\cite{Gull:rmp/83/349,Phillips:jcp/140/241101,Kananenka:prb/91/121111,:/content/aip/journal/jcp/142/19/10.1063/1.4921259}.

\begin{figure}[htbp]
\begin{center}
\includegraphics[width=0.49\textwidth]{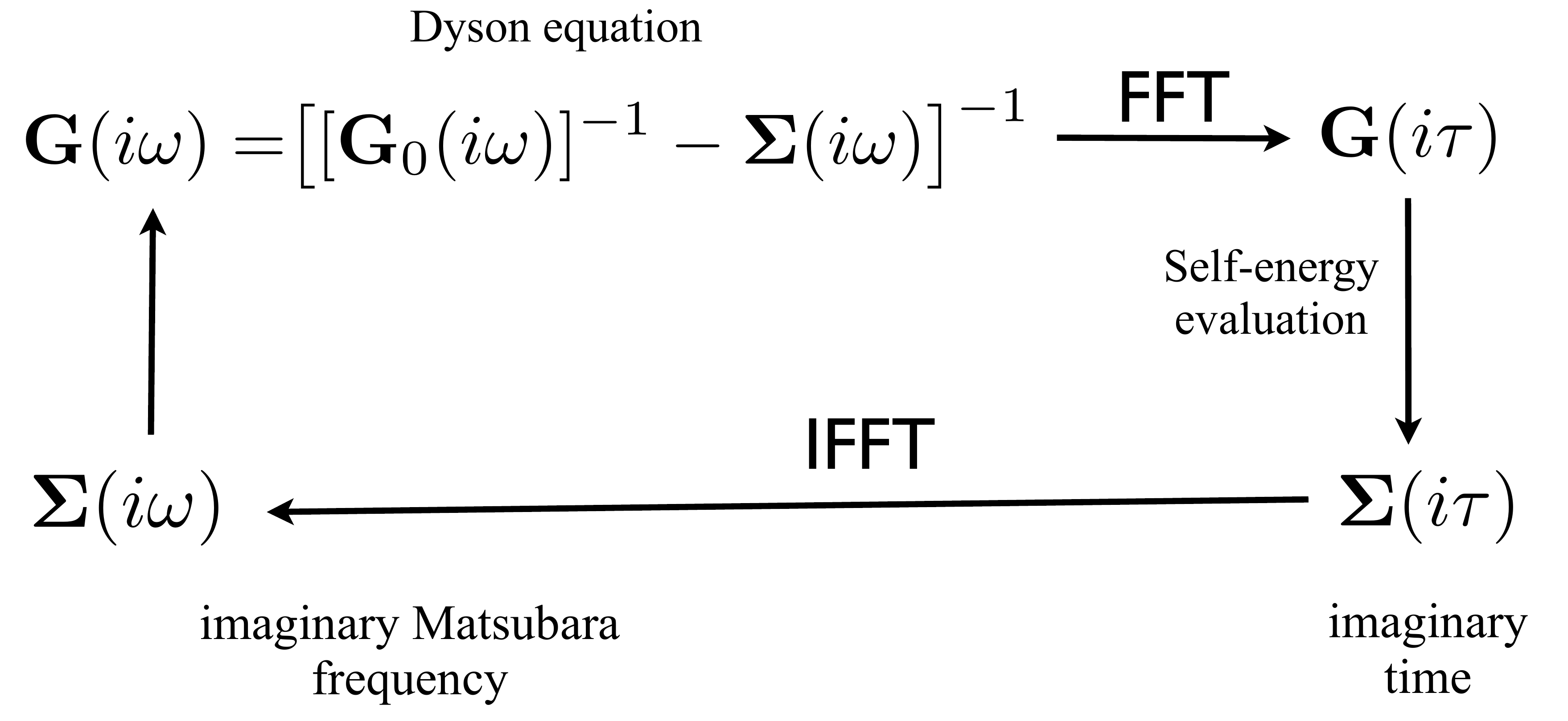}
\caption{An example of the representation change between 
imaginary time and frequency in Green's functions methods. 
We denote the non-interacting and interacting/correlated Green's 
functions as $G_0(i\omega)$ and $G(i\omega)$, respectively.}
\label{fig:ft_scheme}
\end{center}
\end{figure}

Thus, frequently a Green's function method proceeds according to 
scheme illustrated in Figure~\ref{fig:ft_scheme} and the computational 
bottleneck lies in the evaluation of $\Sigma(i\tau)$. 

The number of imaginary time points used to represent $\Sigma(i\tau)$ enters as a prefactor. 
However, a sizable prefactor makes such calculations for large 
system impossible.

There are two reasons for the large size of imaginary time grids 
for temperature-dependent Green's functions, firstly enough 
points are required to preserve the temperature dependence, 
secondly a significant number of points is required to make a 
numerical Fourier transform accurate. 

\subsection{Orthogonal polynomial representation of self-energy}

We consider a single-particle temperature-dependent imaginary time
self-energy $\Sigma(i\tau)$ on the interval $\left[0,\beta\right]$.
The self-energy can be expanded in Legendre--Fourier series using 
orthogonal polynomials $P_l$. 
Note that different orthogonal polynomials, e.g. Chebyshev or Legendre,
can be used as a basis. However, all results presented here should be 
valid regardless of the representation~\cite{Boehnke:prb/84/075145,li:arxiv/1205.2791}.
In this work, we use Legendre polynomials defined on the interval 
$\left[-1,1\right]$. The Legendre polynomial expansion of the self-energy 
on this interval is given by
\begin{equation}
\Sigma_{ij}(i\tau) = \sum_{l \ge 0}^\infty \frac{\sqrt{2l+1}}{\beta} P_l(x(i\tau)) \Sigma^l_{ij},
\end{equation}
where $P_l(x(i\tau))$ is a Legendre polynomial of rank $l$ and 
$\Sigma^l_{ij}$ is the corresponding expansion coefficient,
\begin{equation}
 \Sigma^l_{ij} = \sqrt{2l+1} \int_0^\beta d\tau P_l(x(\tau)) 
\Sigma_{ij}(i\tau),
 \label{eq:coef}
\end{equation}
and $x(\tau)=2\tau/\beta-1$ maps interval 
$[0,\beta]$ onto $\left[-1,1\right]$.
For detailed description of the properties of Legendre polynomials 
see ref~\citenum{abramowitz+stegun}.

In the past, several advantages of using an orthogonal polynomial 
representation of Green's function (and self-energy) were explored. 
Boehnke et. al.~\cite{Boehnke:prb/84/075145} used the Legendre
polynomial representation of Green's functions
as a noise filter in continuous-time hybridization expansion quantum 
Monte Carlo (CT-HYB) calculations. 
Additionally, the Legendre representation of Green's function can be 
more compact than Matsubara Green's function~\cite{Boehnke:prb/84/075145,Arsenault:prb/90/155136,wu:cpl/30/090201}.

In this paper, we describe another advantage of using Legendre 
representation. It allows us to perform an orthogonal polynomial two 
step transform ($\Sigma(i\tau) \rightarrow L$) and 
($L \rightarrow \Sigma(i\omega)$) and to reduce the size of the 
imaginary time grid that would be otherwise necessary to maintain a 
high accuracy.
This ($\Sigma(i\tau) \leftrightarrow L$) transformation is unitary and 
can be written as a matrix multiplication
\begin{align}
\Sigma_{ij}(i\omega_n) = & \sum_{l \ge 0}^\infty \Sigma^l_{ij}  \frac{\sqrt{2l+1}}{\beta} \int_0^\beta d\tau
e^{i\omega_n \tau} P_l(x(\tau)) \nonumber \\
= & \sum_{l \ge 0}^\infty T_{nl} \Sigma^l_{ij},
\end{align}
where $T_{nl}$ is the unitary matrix with elements defined as
\begin{equation}
T_{nl} = (-1)^n i^{l+1} \sqrt{2l+1} j_l \left(
\frac{\left(2n+1 \right)\pi}{2}
\right),
\label{eq:ft}
\end{equation}
where $j_l(z)$ are the spherical Bessel functions of the second 
kind.

Even though the Legendre series is infinite in principle, in all 
practical calculations only a finite number of expansion coefficients
is used. For all atomic and molecular systems studied here the 
expansion coefficients decay very fast. In the worst case, only a 
few hundreds of them are necessary (see section~\ref{results}),
therefore only $O(N_l n^2)$ double precision numbers 
have to be stored, where $n$ is the number of orbitals and $N_l$ is the
number of terms in the Legendre expansion. In contrast to calculations 
employing orthogonal polynomial transform, a typical 
numerical Fourier transform may require tens to hundreds of thousands 
of imaginary time grid points $N_\tau$, making 
the cost of evaluation of the self-energy very significant.

\subsection{Sparse imaginary time grid}

In this section, we examine the number of imaginary time grid points 
necessary to perform the Fourier transform ($\Sigma(i\tau) \rightarrow L$) 
accurately. 
This assessment is absolutely vital to the success of many approaches 
where evaluation of the self-energy on the time grid is the computation 
bottleneck. 

To estimate the number of grid points necessary we performed 
calculations for atoms and molecules using different numbers of 
imaginary time points.
While both atoms and molecules that we use here as test examples do not display 
different physics for a large range of temperatures due to the size of the gap 
present in these systems, they are very challenging examples since we strive to 
calculate electronic energy for very low temperatures near absolute zero
(large value of inverse temperature $\beta=100$). Consequently, our grid 
spacing has to be very small and the grid has to span significant energy window 
shown in Figure~\ref{fig:energy_window} thus requiring very many points if the 
uniform or power-law grids are used.
For numerical Fourier transforms we used two different grids, 
a uniform grid and power-law grid as described below. For the 
($\Sigma(i\tau) \rightarrow L$) transform, we used a modification 
of the power-law grid to compute a fixed number of Legendre coefficients.
We have chosen 200 expansion coefficients since such a number of 
coefficients allows us to calculate the energy to $\mu$Ha accuracy.


\subsubsection{Uniform imaginary time grid}
The grid points are uniformly spaced within the interval $[0,\beta]$ 
where the antiperiodic $G(i\tau)$ or $\Sigma(i\tau)$ is represented.

\subsubsection{Power-law time grid}
Since imaginary time self-energy is sharply peaked around endpoints 
(0 and $\pm \beta$) it is convenient to use non-uniformly 
spaced grids to represent it. A power-law grid~\cite{ku:prl/89/126401}
is constructed to be dense around endpoints and sparse between them 
where imaginary time self-energy is close to zero. 
The power-law grid is defined by two parameters: the power coefficient 
$p$ and the uniform coefficient $u$. The first step in creating such a 
grid is placing points with the coordinates $\tau_j=\beta/2^j, j\in\{0,...,p-1\}$ starting from each endpoint
and also placing a midpoint at $\beta/2$. 
Consequently, such a grid has $2p+1$ points.
Then, each interval between power points is divided into $2u$ 
uniformly spaced subintervals. 

\subsubsection{($\Sigma(i\tau) \rightarrow L$) transform with a power-law grid}\label{sec:aft_pg}

We evaluated Legendre expansion coefficients employing only a fraction 
of the original power-law grid. We kept number of power points fixed $p=12$ and 
choose the number of uniform points as $u=2^n$ where $n \in \{0,...,5\}$.
The resulting grid has $2u(p+1)+1$ imaginary time points that correspond 
to 53, 105, 209, 417 and 833 points for $n \in \{0,...,5\}$ respectively. 

\begin{figure}[htbp]
\begin{center}
\includegraphics{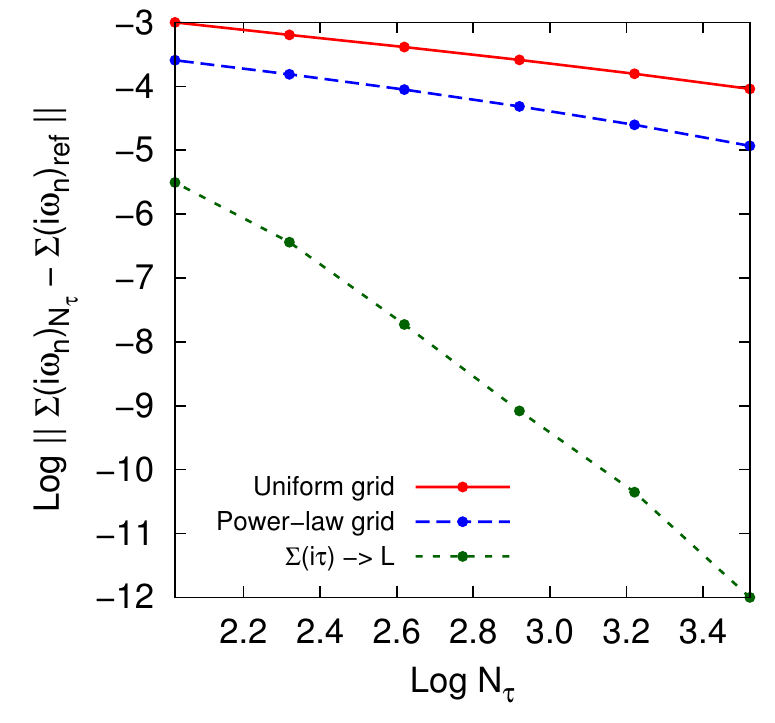}
\caption{Convergence of the self-energy $\Sigma(i\omega_n)$
as a function of number of imaginary time points.}
\label{fig:grid_convergence}
\end{center}
\end{figure}

Using the uniform, power-law grid and orthogonal polynomial transform, we 
evaluated the Matsubara self-energy $\Sigma(i\omega)$. In Figure~\ref{fig:grid_convergence}, 
we show convergence for different grids. We set as our reference the 
self-energy obtained with 200 Legendre polynomials.

As seen in Figure~\ref{fig:grid_convergence}, the  
($\Sigma(i\tau) \rightarrow L$) and ($L \rightarrow \Sigma(i\omega)$) 
transforms converge much faster than the regular 
($\Sigma(i\tau) \rightarrow \Sigma(i\omega)$) Fourier transform performed 
using a uniform or power-law grid.  

\section{Self-consistent second-order Green's function theory 
using orthogonal polynomial transform}\label{GF2}

\begin{figure}[htbp]
\begin{center}
\includegraphics[width=0.49\textwidth]{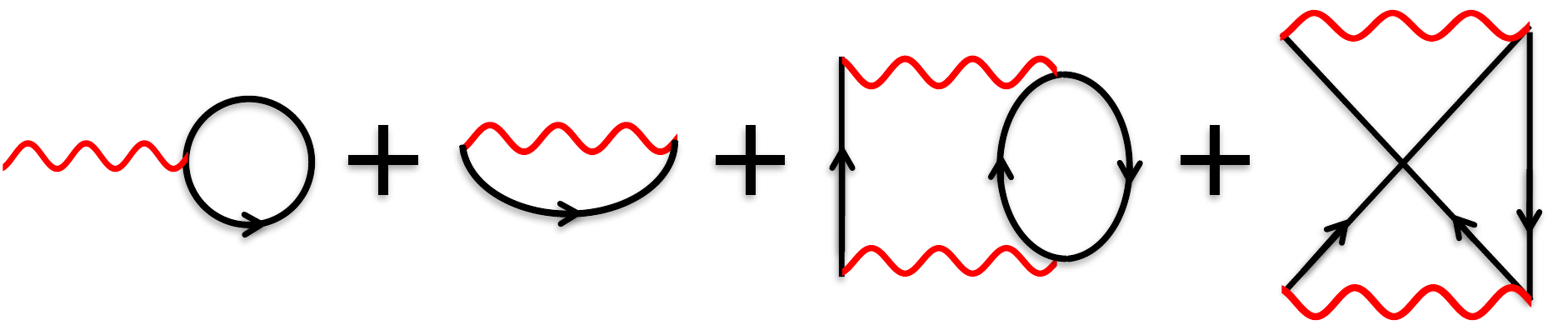}
\caption{GF2 is the second order approximation to the self-energy.
It includes two first order self-energy diagrams (from left to right): 
Hartree (direct), Fock (exchange) and two second order self-energy diagrams: direct and exchange.}
\label{fig:feynmann}
\end{center}
\end{figure}

In this section, as an application to realistic calculations, we briefly 
describe the framework of the self-consistent second-order Green's function 
theory (GF2) using an orthogonal polynomial transform.
 
GF2 employs the second-order approximation to the self-energy resulting 
in inclusion of all Feynman self-energy diagrams up to the second order, 
as shown in Figure~\ref{fig:feynmann}. The first two diagrams are already 
included at mean-field level.  Within spin-restricted GF2, the last two 
diagrams are translated into the following expression for the imaginary time
self-energy
\begin{align}
\Sigma_{ij}(i\tau_n) & =  -\sum_{klmnpq} G^0_{kl}(i\tau_n) G^0_{mn}(i\tau_n) G^0_{pq}(-i\tau_n)\nonumber \\
& \times v_{imqk} \left( 2v_{lpnj} - v_{nplj} \right ),
\label{eq:gf2}
\end{align}
where $G^0(i\tau_n)$ is the zeroth order imaginary time Green's function and $v_{ijkl}$ are two-electron
integrals defined as 
\begin{equation}
v_{ijkl}=\int \int d{\bf r}_1 d{\bf r}_2 \phi^{*}_{i}({\bf r}_1 )\phi_{j}({\bf r}_2)\frac{1}{r_{12}}\phi^{*}_{k}({\bf r}_1)\phi_{l}({\bf r}_2).
\end{equation}
In GF2, as illustrated in Figure~\ref{fig:ft_scheme} for a general case, 
the calculation of the second-order self-energy according to 
eq~\ref{eq:gf2} is done in the imaginary time domain while
the Dyson equation from Figure~\ref{fig:ft_scheme} is much easier to 
solve in the frequency domain. 

In our previous implementation, a typical molecular all electron GF2 
calculation that involves both core and virtual orbitals may require 
many thousands of Matsubara frequencies $N_\omega$ making the total 
amount of storage necessary equal to $O(N_\omega n^2)$, 
where $n$ is the number of orbitals. The imaginary time Green's function 
is represented by $O(N_\tau n^2)$ double precision numbers, 
where $N_\tau$ is the number of points of the imaginary time grid.
Building the self-energy according to eq~\ref{eq:gf2} scales as 
$O(N_\tau n^5)$, and despite that the self-energy calculation 
at any given imaginary time point is independent and can be made parallel, a large 
prefactor $N_\tau$ is slowing down calculations significantly even 
when using a power-law grid.
%
%

Since employing the orthogonal polynomial transform restricts the 
imaginary time grid even for the most difficult cases to fewer than 
400 points, we implemented it as part of our algorithm. 
Here, we give a complete step-by-step modified algorithm description.

\begin{enumerate}

\item Start with Hartree--Fock (HF) reference solution 
(although starting from DFT reference is equally possible and advantageous 
for cases that are difficult to converge using HF) and build initial 
Matsubara Green's function in non-orthogonal AO basis according to:
\begin{equation}
\mathbf{G}_0(i\omega_n) = \left[ (\mu + i\omega_n) \mathbf{S} - \mathbf{F} \right]^{-1},
\end{equation}
where $\mathbf{S}$ is the overlap matrix, $\mathbf{F}$ is the Fock matrix and $\mu$ is the chemical potential.

\item Perform discrete Fourier transform of $\mathbf{G}_0(i\omega_n)$ 
into its imaginary time counterpart $\mathbf{G}_0(i\tau)$. 
Alternatively, at this point, it is possible to avoid discrete Fourier 
transform if one starts directly from the imaginary time HF Green's 
function constructed from HF orbital energies
\begin{equation}\label{G_tau_direct}
G_{ij}^0(i\tau_n) = \theta (i\tau_n) \left( n(E_i) - 1 \right)e^{-E_i i\tau_n}
+ \theta (-i\tau_n) n(E_i)e^{-E_i i\tau_n},
\end{equation}
where $\theta(x)$ is Heaviside step function, $E_i=\epsilon_i-\mu$ 
and $n(E_i)=1/(e^{\beta E_i}+1)$ is the Fermi distribution.
Since the Green's function from eq~\ref{G_tau_direct} is constructed 
using MO orbital energies, it should be transformed to AO basis before 
proceeding to the next step.

\item Calculate the self-energy on the imaginary time grid according to 
eq~\ref{eq:gf2}. It is at this point where we first take advantage 
of Legendre polynomial representation of
self-energy, since the Legendre representation allows us to use small 
imaginary time grids with only a fraction of points of the grid
we used in our original 
implementation~\cite{Phillips:jcp/140/241101}.\label{self_energ_elval}

\item Obtain the Legendre expansion coefficients by performing an 
integration of the self-energy $\mathbf{\Sigma}(i\tau)$ according to 
eq~\ref{eq:coef}.

\item Build the imaginary frequency self-energy $\mathbf{\Sigma}(i\omega)$ 
by performing a transform of Legendre coefficients according to 
eq~\ref{eq:ft}.

\item Solve the Dyson equation to obtain an updated Green's 
function.\label{end_small_loop}

\item Find the chemical potential $\mu$ to ensure that a proper number 
of electrons is present in the system.

\item Calculate the density matrix and use it to update Fock matrix.

\item Go to point \ref{end_small_loop} and iterate until the density 
matrix and chemical potential $\mu$ stop to change.

\item Calculate the one-body energy as 
\begin{equation}
E_{1b}=-\frac{1}{2}\sum_{\mu}\sum_{\nu}P_{\nu\mu}(t_{\mu\nu}+F_{\mu\nu}),
\end{equation}
where $P_{\nu\mu}=-2G_{\nu\mu}(i\tau=\beta)$ is the correlated density 
matrix and the Fock matrix is evaluated using this correlated density matrix.
The two-body energy can be evaluated using as
\begin{equation}
E_{2b} = -\frac{1}{\beta}\sum_{n=0}^{n_{freq}}\tr [G(i\omega_n)\Sigma(i\omega_n)],
\label{eq:ecor}
\end{equation}
for details see ref~\citenum{Fetter}.

\item Transform $G(i\omega)$ to $G(i\tau)$ and go to 
step~\ref{self_energ_elval} and iterate until the total energy converges.

\end{enumerate}

\section{Results and discussion}\label{results}

In this section, we provide results of atomic and molecular 
calculations with the above introduced GF2 algorithm.
To assess the accuracy and efficiency of the algorithm described 
above, we performed several benchmark
calculations with large basis sets using diffuse orbitals. These basis 
sets usually required the most extensive imaginary time grid 
and are necessary to reach quantitative accuracy and converge with 
basis set size. 

Additionally, we also tested our algorithm on a few systems with 
transition metal atoms with 
ecp-sdd-DZ~\cite{Dolg:jcp/86/866,Stoll:jcp/79/5532,JCC:JCC9,Schuchardt:jcif/47/1045} 
basis set containing 
pseudopotentials for inner shell electrons. We investigated 
systems with pseudopotentials since these are frequently used in 
solid state calculations and it is our interest to assess how compact 
the grids can become for such systems. 

Our investigations can be divided into two groups evoking our original 
questions about {\bf(i)} the size of the grid necessary to calculate 
the Legendre coefficients accurately and 
{\bf(ii)} the compactness of the Legendre expansion. 

\begin{figure}[htbp]
\begin{center}
\includegraphics{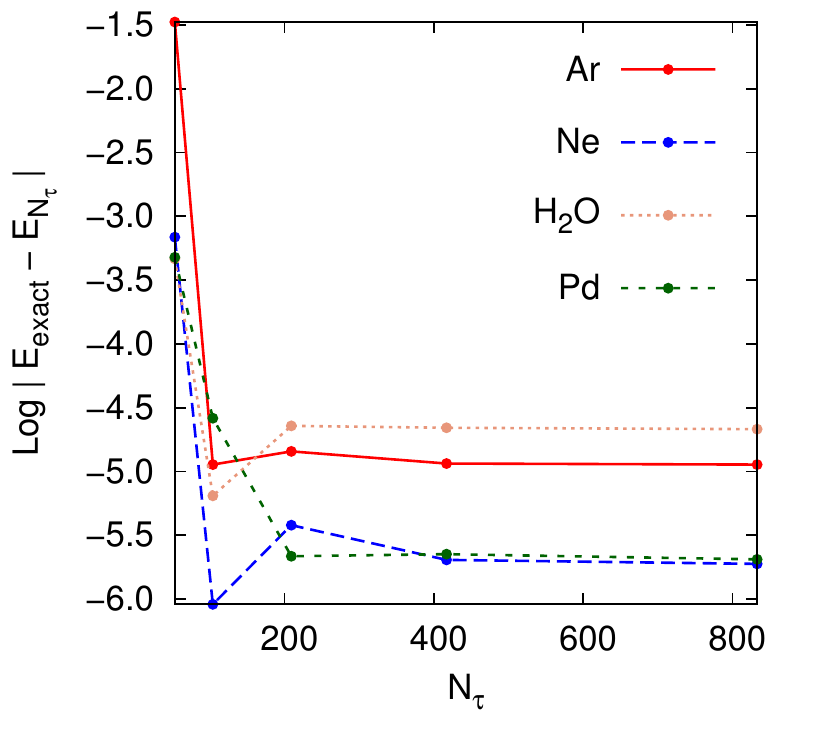}
\caption{Convergence of electron correlation energy with the size of 
imaginary time grid $N_\tau$
calculated for a few atoms and molecules.}
\label{fig:grid}
\end{center}
\end{figure}

\begin{figure*}[htbp]
\begin{center}
\includegraphics{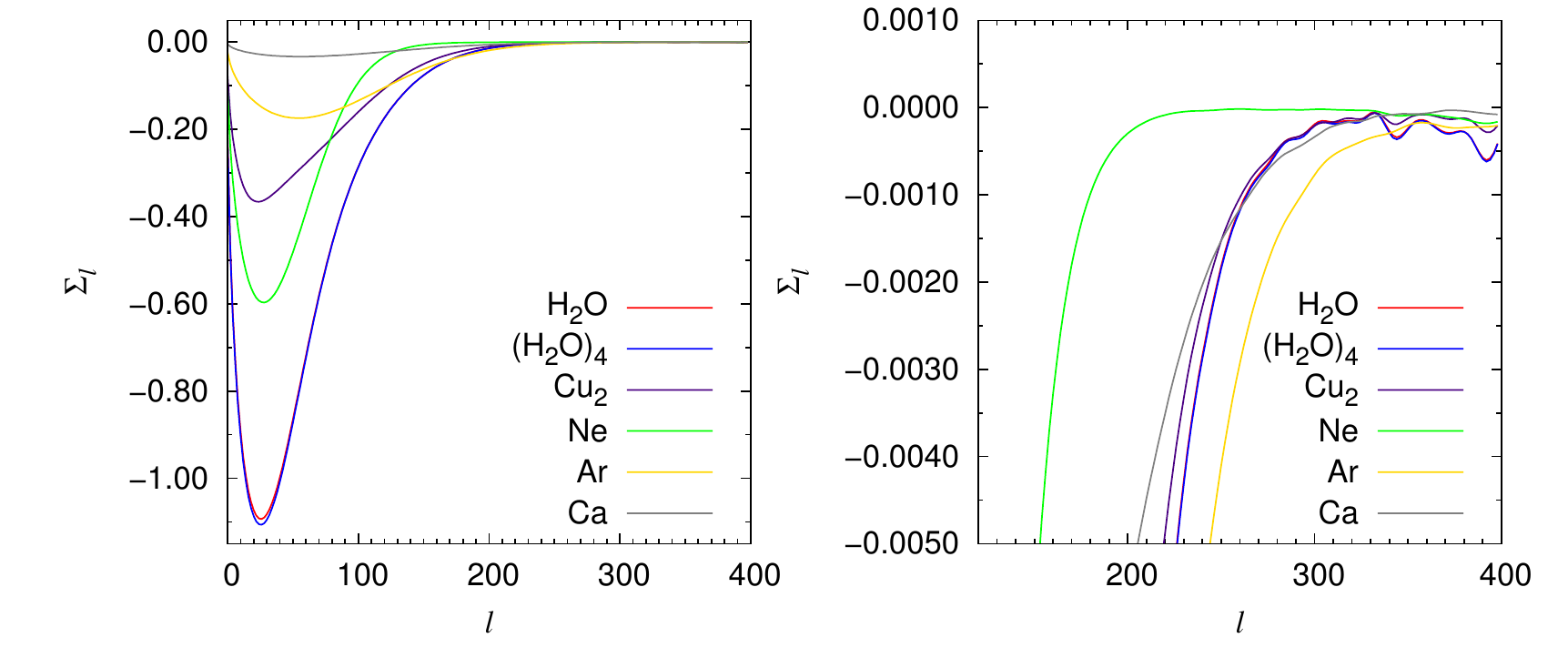}
\caption{Even coefficients of Legendre expansion of self-energy 
calculated using imaginary time grid consisting of
417 points for atoms and molecules. 
Left panel shows zoomed in region where $\Sigma^l \approx 0$ where 
Gibbs oscillations start to develop.}
\label{fig:coef_mol}
\end{center}
\end{figure*}

\subsection{Convergence of the electronic energy with respect to the grid size}

We calculated the electronic correlation energy according to 
eq~\ref{eq:ecor} for the grids defined in Section~\ref{sec:aft_pg} 
and compared it to the one obtained using our previous GF2 implementation
with sufficiently large imaginary time grids. Our previous implementation 
required at least an order of magnitude larger grids than the current 
one using the Legendre expansion. 

In Figure~\ref{fig:grid}, we plot the error in
electronic correlation energy obtained by using different number of 
imaginary time points that are used to produce Legendre expansion of 
self-energy consisting of 200 expansion coefficients. 
Let us first note that using 200 Legendre coefficients results on average 
in 20 or less $\mu$Ha error in the total energy, see 
Table~\ref{tab:enrg_err200}. 
\begin{table}
  \caption{Error in $\mu$Ha of the total energy for different number of 
   Legendre expansion coefficients for several atoms and molecules.}
   \label{tab:enrg_err200}
    \begin{ruledtabular}
\begin{tabular}{c | r | r |  r}
 atom or molecule & l=200 & l=100 & l=40 \\
\hline
   He                                      &  0.09     &  0.05     &   -44.47   \\
   Be                                      &  0.30     &  0.02     &   -8.29    \\
   Ne                                      &  -1.05    &  -69.57   &   -177.23  \\
   Mg                                      &  0.50     &  -6.70    &   -6.28    \\
   Ar                                      &  -9.93    &  -28.53   &   697.07   \\
   Ca                                      &  -3.42    &  -7.54    &   401.17   \\
   H$_2$O\textsuperscript{\emph{a}}        &  -22.00   &  -99.99   &    9478.18 \\
   (H$_2$O)$_2$\textsuperscript{\emph{a}}  &  -27.26   &  372.33   &    21048.60\\
   (H$_2$O)$_3$\textsuperscript{\emph{a}}  &  -53.20   &  47.81    &    29737.13\\
   (H$_2$O)$_4$\textsuperscript{\emph{a}}  &  -71.21   &  63.47    &    39668.95\\
   NH$_3$\textsuperscript{\emph{b}}        &  -7.12    &  -225.17  &    6953.89 \\
   C$_6$H$_6$\textsuperscript{\emph{b}}    &  -7.87    &  -668.08  &    28778.84\\
   Cd                                      &  -2.27    &  -38.27   &    201.04  \\
   Cu$_2$\textsuperscript{\emph{c}}        &  -31.09   &  -151.26  &    6794.16 \\
   Pd                                      &  -2.25    &  -48.21   &    121.04  \\
   Ag$_2$\textsuperscript{\emph{d}}        &  -4.67    &  -36.86   &    658.88  \\
  \end{tabular}
\end{ruledtabular}
  \textsuperscript{\emph{a}} Geometry was taken from ref~\citenum{Wales199865}.\\
  \textsuperscript{\emph{b}} Experimental geometries were taken from NIST Computational Chemistry Comparison and Benchmark Database~\cite{nist}.\\
  \textsuperscript{\emph{c}} $d$(Cu-Cu)=3.63 a. u.\\
  \textsuperscript{\emph{d}} $d$(Ag-Ag)=5.46 a. u.\\
\end{table}

If we set the value of correlation energy using 200 Legendre polynomials as a 
reference then it is evident from Figure~\ref{fig:grid}
that with only 104 imaginary time
grid points an acceptable accuracy (less than 50 $\mu$Ha from the 
exact answer) can be achieved. This is already far below the commonly 
accepted chemical accuracy of $\approx 1$ kcal/mol. Using fewer 
than 100 grid points is not advisable since for 53 points for every 
examined case the accuracy was about 0.5 $m$Ha away from the exact 
answer that is unacceptable in almost any quantum chemical calculation. 
For imaginary time grids with 209 and more points the accuracy reaches 
a plateau since for 200 Legendre coefficients a grid of 209 frequency 
points is sufficient to produce these coefficients with accuracy 
reaching numerical precision. 
If we desired to reach better accuracy than $\mu$Ha level, then a 
larger number of the Legendre expansion coefficients should be 
employed in our calculation.

\begin{figure}[htbp]
\begin{center}
\includegraphics{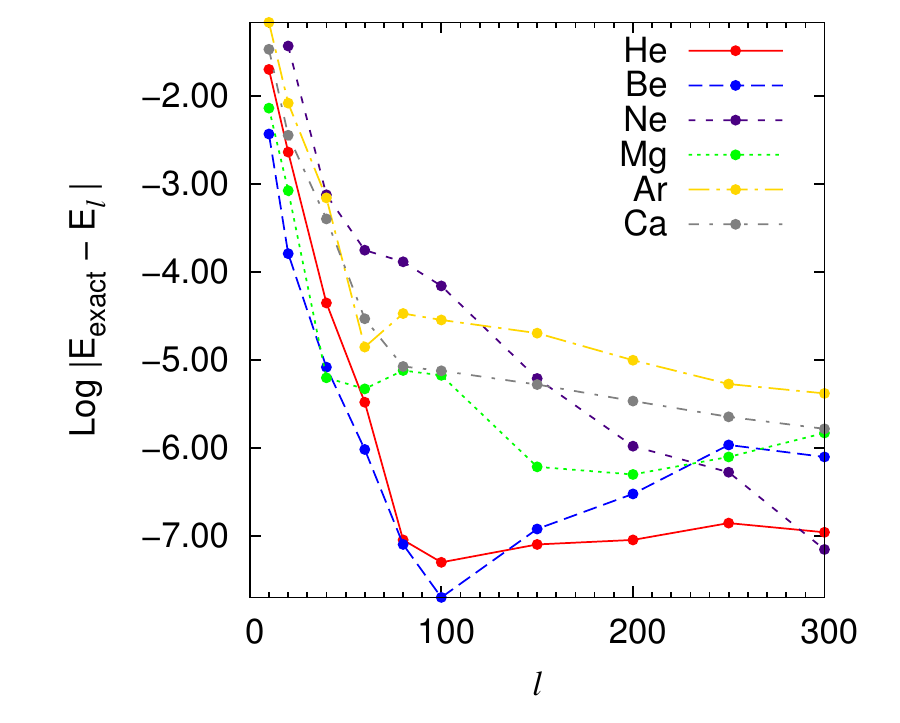}
\caption{Difference between the standard GF2 correlation energy and 
correlation energy obtained by using Legendre expansion of the 
self-energy for several closed-shell atoms with aug-cc-pVDZ 
basis set (cc-pVDZ basis set was used for Ca).}
\label{fig:cvg_atoms}
\end{center}
\end{figure}

\begin{figure}[htbp]
\begin{center}
\includegraphics{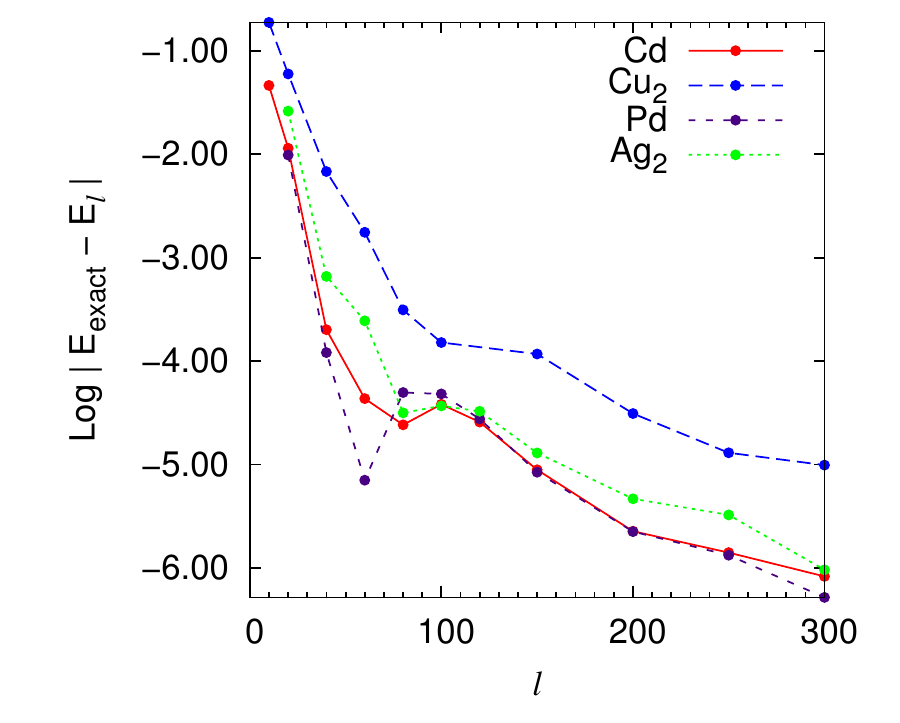}
\caption{Difference between the standard GF2 correlation energy 
and correlation energy obtained by using Legendre expansion of 
the self-energy for several transition metal atoms and diatomic 
clusters with ecp-sdd-DZ basis set.}
\label{fig:cvg_pseudo}
\end{center}
\end{figure}

\begin{figure}[htbp]
\begin{center}
\includegraphics{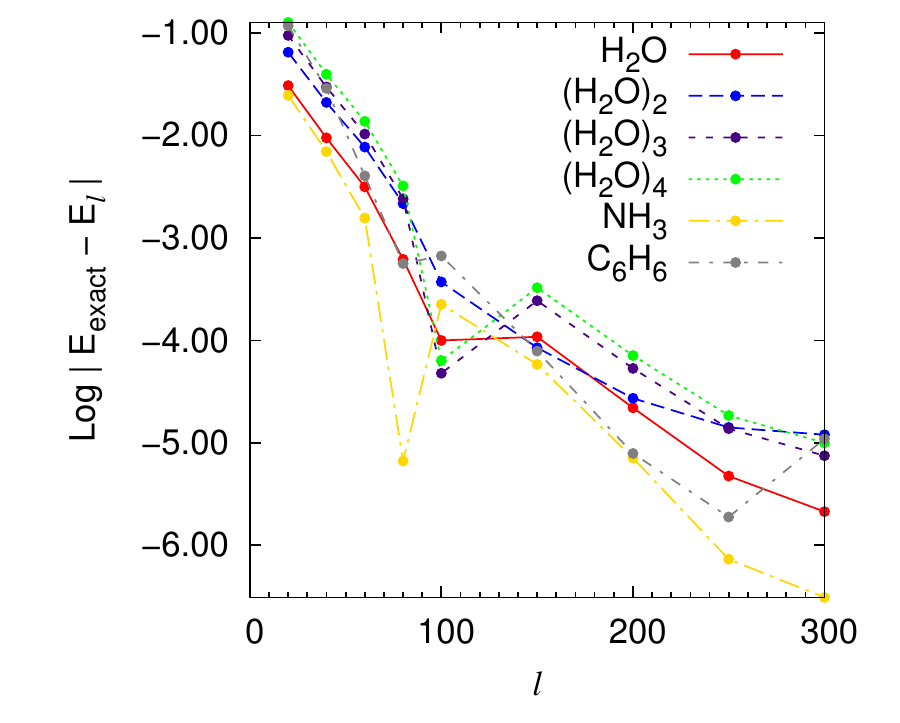}
\caption{Difference between the standard GF2 correlation energy 
and correlation energy obtained by using Legendre expansion of 
the self-energy for several small molecules with TZ basis set.}
\label{fig:cvg_mol}
\end{center}
\end{figure}

\subsection{Compactness of the Legendre expansion}

Next, we will study how the accuracy of our calculations depends 
on the number of terms in Legendre expansion of the self-energy. 
We fix the size of imaginary time grid to 417 points because, as it
is seen from Figure~\ref{fig:grid}, this number of imaginary frequencies 
Is sufficient to produce at least 200 accurate Legendre coefficients in 
the expansion of the self-energy.

First, we will look at values of expansion coefficients $\Sigma_{00}^l$ 
determined by integrating self-energy according to eq~\ref{eq:coef}.
In Figure~\ref{fig:coef_mol} we plotted even Legendre coefficients for a 
few atoms and molecules. Odd coefficients show very similar behavior 
in almost all cases.
As seen from the left panel of Figure~\ref{fig:coef_mol}, Legendre 
coefficients decay monotonically, converging to zero with the decay 
rate that is system-specific. For all cases studied in this work, 
we observe a fast decay of Legendre expansion coefficients used to 
represent the imaginary time self-energy. Thus, the Legendre polynomials 
form a compact representation not only for
Hubbard model~\cite{Boehnke:prb/84/075145} but also for atoms and molecules. Since realistic molecular systems have diffuse orbitals and span large 
energy spectrum an increase in the number of expansion terms in 
comparison to the Hubbard model is to be expected.

The right panel of Figure~\ref{fig:coef_mol} shows a zoom in the region 
where the expansion coefficients are close to zero,  $\Sigma^l_{00}=0$.
A closer look at the values of $\Sigma^l_{00}$ reveals a numerical noise. 
This noise is known as Gibbs's oscillations~\cite{Weisse:rmp/78/275} and 
arises when too few imaginary time grid 
points are used to evaluate higher order Legendre coefficients.
In order to prevent numerical noise buildup affecting very high orders 
of the Legendre expansion, one should truncate the expansion once 
oscillations are detected. 
Another method is to damp Gibbs oscillations by introducing an integral 
kernel function~\cite{Weisse:rmp/78/275}. This option has been previously 
explored in the context of the Hubbard model~\cite{li:arxiv/1205.2791}. 
The particular choice of integral kernel function depends on several 
factors and is not convenient, especially if a black-box method is desired. 
Motivated by our aspiration to keep the self-energy as compact as possible,
we took the first route and studied how the truncation of the Legendre 
expansion of the self-energy influences the accuracy 
of the method. The truncation of the expansion is unambiguous and the 
resulting self-energy is free from Gibbs oscillations.
It is also worth mentioning that truncation criteria can be easily 
implemented and do not require any special care from the user
and hence can be introduced as a part of any black-box computational 
package.

We performed calculations for our test set containing atoms and simple 
molecules truncating the Legendre series after various number of 
terms $\Sigma^l_{ij}=0$ for $l>l_\text{cut}$.  

Figures~\ref{fig:cvg_atoms}-\ref{fig:cvg_mol} show that only couple 
hundreds of Legendre expansion coefficients are necessary to yield 
very accurate results. All these results were calculated using 417 
imaginary time points. In all cases very fast convergence was achieved
and less then 100 Legendre polynomials were needed to converge 
correlation energy to 1 $m$Ha. 
We observe that for all atoms considered here the energy
continues to converge to the exact one and about 200 Legendre polynomials 
are needed to recover it up to the $\mu$Ha from the exact result. 
Similar observations can be made for calculations involving 
pseudopotentials and thus our method can be reliably applied to 
calculations of complex systems containing both transition metals and 
light atoms. A somewhat slower convergence was observed in the case 
of molecular calculations. In this case more than 200 but less than 
300 Legendre polynomials were necessary to get within few $\mu$Ha 
from the exact energy.


\section{Summary and Conclusions}\label{conclusions}

The frequency---time duality is present in many temperature-dependent 
methods and $G(i\tau)$ or $\Sigma(i\tau)$ can be transformed to 
$G(i\omega)$ or $\Sigma(i\omega)$ depending if handling the time or 
frequency object is more computationally advantageous. 
Both the frequency and time Green's functions have to be represented 
on a numerical grid, thus a small number of grid points is crucial 
for achieving computational efficiency.

While the construction of imaginary time grids is a well-studied problem 
and appears in Laplace-transformed M\o ller--Plesset (LT-MP2) perturbation 
theory, such grids are appropriate for the zero-temperature Green's 
function but cannot be employed to study the temperature-dependent 
Green's function. 

In this paper, we have presented a method that makes transform between 
imaginary time and imaginary frequency for temperature-dependent 
Green's function converge much faster with respect to the number of 
necessary imaginary time points than the traditional uniform and 
power-law grid. To achieve this goal we have used a combination of a very 
sparse power-law grid together with explicit transform based on a Legendre 
expansion of the self-energy. We have shown that to 
converge the Legendre coefficients necessary to perform the explicit 
transform we need an order of magnitude fewer imaginary time points 
that when doing the numerical Fourier transform.
 
Moreover, we have also shown that even for realistic systems in 
basis sets with a significant energy spread only a few hundred (200-300) 
of Legendre coefficients are necessary to reach the accuracy of $\mu$Ha 
when compared with the fully converged result. Overall the orthogonal 
polynomial representation of self-energy offers fast, accurate, and less 
storage demanding solution for practical realistic calculations.

We have also applied such a representation of the self-energy to the GF2 
method resulting in very accurate energies for atoms and molecules while 
using only a limited number of imaginary time grid points and only couple hundred 
(200-300) of Legendre expansion coefficients.

While at present no large scale realistic calculations are performed 
including temperature coming from the electronic effects, due to the 
increasing interest in new materials, we believe that such calculations 
will become important in the near future and identifying and overcoming 
major bottlenecks connected to the efficient representation of Green's 
function and self-energy in the imaginary time domain is an important 
step in this new direction.

\section*{Acknowledgments}
A. A. K., J. J. P, and D. Z. acknowledge support by DOE ER16391.


%

\end{document}